\documentclass[anonymous=false, %
               format=acmsmall, %
               review=false, %
               screen=true, %
               nonacm=true]{acmart}

\usepackage[utf8]{inputenc} %
\usepackage[T1]{fontenc}    %
\usepackage{hyperref}       %
\usepackage{url}            %
\usepackage{booktabs}       %
\usepackage{amsfonts}       %
\usepackage{nicefrac}       %
\usepackage{microtype}      %
\usepackage{amsmath}
\usepackage{graphicx}
\usepackage{listings}
\usepackage{xcolor}
\usepackage{courier}
\usepackage[nameinlink]{cleveref}
\creflabelformat{equation}{#2\textup{#1}#3}  %

\author{Pavel Sountsov}
\email{siege@google.com}
\author{Alexey Radul}
\email{axch@google.com}
\author{Srinivas Vasudevan}
\affiliation{ Google Research}
\email{axch@google.com}
\title{FunMC: A functional API for building Markov Chains}
\begin{document}

\begin{abstract}
Constant-memory algorithms, also loosely called Markov chains, power
the vast majority of probabilistic inference and machine learning
applications today.
A lot of progress has been made in constructing user-friendly APIs
around these algorithms.
Such APIs, however, rarely make it easy to research new
algorithms of this type.
In this work we present FunMC, a minimal Python library for
doing methodological research into algorithms based on Markov
chains. 
FunMC is not targeted toward data scientists or others who
wish to use MCMC or optimization as a black box, but rather towards
researchers implementing new Markovian algorithms from scratch.
\end{abstract}

\maketitle

\section{Introduction}

In machine
learning, gradient descent is a Markov chain of the parameters of the model
being optimized, often augmented with some auxiliary quantities for methods
such as Adam \citep{kingma2015adam}. In probabilistic inference, Markov chain
Monte Carlo (MCMC) is a Markov chain of the values of random variables, also
often augmented with auxiliary variables \citep[e.g.,][]{neal2011mcmc}.
Such algorithms are pervasive in the practice of machine learning,
probabilistic and otherwise.  As such, they have received a great deal of
well-deserved attention among library and API developers
\citep[e.g.,][]{salvatier2016probabilistic,bingham2018pyro,tfp}.

We present FunMC, a Markov chain library designed specifically to
enable and accelerate research into new constant-memory algorithms for
machine learning and probabilistic inference.  The design of FunMC
follows these principles:

\begin{enumerate}

  \item All API elements are completely stateless, simplifying
    reasoning, composition, and interoperation with function-oriented
    platforms like JAX \citep{jax2018github} and subsystems like
    automatic differentation (through the Markov chain).

  \item Pervasive support for returning and propagating side information.

  \item Composition over configuration. All API elements aim to do one thing
      well, eschewing long argument or flag lists where possible.

  \item A unified API for MCMC, optimization, and running
    statistics, for construction of hybrid algorithms.

\end{enumerate}

We have implemented FunMC as a multi-backend library of Python
functions, executable with TensorFlow \citep{tensorflow} and JAX.
It should be relatively easy to port to any other Python machine
learning framework. The code is available at
\url{https://github.com/tensorflow/probability/tree/master/spinoffs/fun_mc}.

The contributions of this work are:

\begin{enumerate}

  \item We believe FunMC is the first framework for Markovian
    algorithms whose components are small enough and simple enough to
    be reused often, while also composing smoothly enough to build
    sophisticated algorithms out of.

  \item We support this claim by demonstrating how to use FunMC for
    MCMC sampling using canned (\Cref{sec:running-example}) or custom
    (\Cref{sec:mcmc}) methods, including thinning (\Cref{sec:thinning}); how to
    reparameterize the target (\Cref{sec:reparameterization}); how to
    do optimization (\Cref{sec:optimization}), including adapting
    parameters like step sizes (\Cref{fig:adaptation}); and how to
    compute streaming statistics (\Cref{sec:streaming-statistics});
    all as independent composable components.

\end{enumerate}

\section{Related work}
\label{sec:related-work}

The FunMC library was designed in response to a number of pre-existing MCMC and
optimization frameworks. Pyro \citep{bingham2018pyro} and PyMC3
\citep{salvatier2016probabilistic} separate the choice of MCMC transition
kernel and the outer sampling loop. The optimization packages of PyTorch
\citep{steiner2019pytorch} and TensorFlow make a similar decomposition between
the loss on one side and the optimization algorithms or outer training loop on
the other. All define a large number of standard MCMC and optimization
transition kernels with arguments to specify the hyperparameters and how to
adapt them. In the case of Pyro and PyMC3, the choice of the adaptation
algorithms is limited, and experimentation with them would require writing new
transition kernels with little or no reuse of existing code. In the case of
PyTorch and TensorFlow, there is support for flexible learning rate adaptation,
but other hyperparameters do not have the same support.

The TensorFlow Probability MCMC (\lstinline{tfp.mcmc}) library \citep{tfp} makes the same factorization of outer
sampling loop and transition kernel, but uses a flexible transition kernel DSL,
where hyperparameter adaptation is effected by nesting transition kernels. For
example, to adapt the step size for Hamiltonian Monte Carlo \citep{neal2011mcmc}, one can wrap a step size
adaptation kernel around
an HMC kernel. A shortcoming of this approach is that it requires complex
message-passing logic between the kernels, forces a DAG structure on the compound
transition kernel computation, and requires learning a new DSL. FunMC does not
distinguish between outer loops and transition kernels, and uses a
more decoupled approach for hyperparameter adaptation, described below.

Modern optimization packages and \lstinline{tfp.mcmc} support definining
losses and target densities as simple callables, using automatic differentiation to compute any
necessary gradients. These callables return the quantity of interest,
(loss or log density), but have limited to no support of returning side
information (e.g., intermediate neural net layer activations). PyMC3 has a
concept of deterministic variables which enables this side channel, but it
requires a DSL for defining probabilistic models. FunMC retains the
flexibility of accepting a simple callable, and makes a simple modification to its
calling convention to support returning side information as per PyMC3.

Pyro and PyMC3 implement reparameterization by taking transport maps as
arguments to their outer sampling loops. TensorFlow Probability implements this
via its transition kernel DSL, where a reparameterization kernel alters the
space on which the wrapped kernels operate on. All of thse approaches limit the
opportunity to adapt the parameters of the transport maps as part of the
sampling procedure. FunMC takes reparameterization completely out of the
transition kernels, using a function transformation instead.

Accelerators based on SIMD operations are at the forefront of high-performance computing in
machine learning today. The \lstinline{tfp.mcmc} library has pioneered using
this capability to run multiple independent chains. Relatedly, NumPyro \citep{phan2019composable}
achieved a similar capability using function transformations. FunMC takes a
hybrid approach, mixing both explicit and automatic batching to run independent
operations in parallel.

\section{A running example: generating samples using MCMC}
\label{sec:running-example}

\begin{figure}[!tb]
\begin{lstlisting}[language=Python, escapechar=|]
def target(w): |\label{line:tlp}|
  logits = input_features @ w
  log_prob = normal_log_prob(w).sum(-1)
  log_prob += bernoulli_log_prob(outputs, logits[..., 0]).sum(-1)
  return log_prob, logits

def kernel(hmc_state, key): |\label{line:kernel}|
  hmc_key, key = jax.random.split(key)
  hmc_state, hmc_extra = fun_mc.hamiltonian_monte_carlo_step(hmc_state,
    target, step_size, num_integrator_steps, seed=hmc_key)
  w, logits = hmc_state.state, hmc_state.state_extra
  return (hmc_state, key), (w, logits, hmc_extra.is_accepted)

(fin_hmc_state, _), (w_chain, logits_chain, is_accepted_chain) = fun_mc.trace(  |\label{line:trace}|
  (fun_mc.hamiltonian_monte_carlo_init(w_init, target), jax.random.PRNGKey(0)),
  kernel, num_steps)
\end{lstlisting}
\caption{Example: HMC on a Bayesian Logistic Regression Model.  Note
  the handling of side information: \lstinline{logits} are returned
  from the target density, combined with the M-H acceptance bit from
  HMC, and their history returned to the user.}
\label{fig:hmc_example}
\end{figure}

As a running example of the FunMC API, consider using Hamiltonian Monte Carlo
(HMC) to sample from the posterior of a simple Bayesian logistic regression
model (\Cref{fig:hmc_example}). We first define the target
density, which is a function with type $\mathtt{State} \rightarrow
(\mathtt{Tensor} , \mathtt{Extra}_l)$, on \Cref{line:tlp}. It is a function
from the value of the random variable $w$ to two return values. The first is
the un-normalized log density evaluated at $w$. The second is arbitrary,
and can be used to return auxiliary information for debugging and other
purposes. In this example we return \lstinline{logits}, which is a function of $w$.

The HMC transition kernel is a function of type $\mathtt{HMCState} \rightarrow
(\mathtt{HMCState}, \mathtt{Extra}_\tau)$, on \Cref{line:kernel}. It conforms
to the concept of a Markov transition kernel by taking and returning the state
of the HMC sampler, but atypically it also has a second output. This second
output is used for returning auxiliary information that is not a
part of the state of the Markov chain. The kernel is a wrapper around the
\lstinline{fun_mc.hamiltonian_monte_carlo_step} transition kernel provided by FunMC with
type $\mathtt{HMCState} \rightarrow (\mathtt{HMCState}, \mathtt{HMCExtra})$.  The wrapper's purpose is to partially apply
\lstinline{fun_mc.hamiltonian_monte_carlo_step}, and to manipulate the side information.

Finally, we iterate our transition kernel on \Cref{line:trace} using the
\lstinline{fun_mc.trace} transition kernel, which has type
$(\mathtt{State}, \mathtt{State} \rightarrow (\mathtt{State}, \mathtt{Extra}), \mathtt{int}) \rightarrow (\mathtt{State},
\mathtt{Extra})$.  This kernel iterates the
function passed as its second argument and stacks its corresponding auxiliary
outputs. We defined \lstinline{kernel} to return the value of $w$, $logits$, and
whether the state proposed by HMC was accepted, which last is available from the
auxiliary return value of \lstinline{fun_mc.hamiltonian_monte_carlo_step}.  The entire history
of this side information is thus available for inspection after the completed
HMC run.

\section{Custom MCMC}
\label{sec:mcmc}

FunMC provides the pre-canned \lstinline{fun_mc.hamiltonian_monte_carlo_step}
kernel for the sake of convenience.  Following the
philosphy of composition over configuration, however, that kernel just consists
(\Cref{sec:appendix}, \Cref{fig:provided_hmc}) of sampling the
auxiliary momentum, running the integrator, and then performing an
accept-reject step, all of which are public functions of FunMC.  This
makes it straightforward for the user to write their own
version if necessary, with maximal code reuse.

\section{\lstinline{fun_mc.trace} transition kernel}
\label{sec:thinning}

The primary usage of the \lstinline{fun_mc.trace} operator is as shown in
\Cref{fig:hmc_example}---computing a (top-level) trace of Markov chain
states.  However, the type of \lstinline{fun_mc.trace} intentionally admits
partial application to produce a transition kernel.  This can be used,
for example, for thinning, as in \Cref{fig:thinning}.  For this purpose,
\lstinline{fun_mc.trace} accepts a \lstinline{trace_mask} argument, which is
prefix-tree-zipped with the auxiliary return. The boolean leaves in the mask
signify whether the corresponding sub-trees in the auxiliary return get traced
or not. A scalar \lstinline{False} matches the entire tree, causing it to be
propagated without history, which is what we want for thinning.

The concept of the kernel returning auxiliary information shows its
value here, by separating the true state of the Markov chain from
information derived from it, but not propagated.
Without framework provision for side information, this effect
would require either statefully mutated side variables, or augmenting
the Markov chain state with the additional quantities that are simply
discarded before each transition. Such a scheme would conceal the Markovian
structure as well as force the user to come up with some initial state for
these auxiliary quntities.

\begin{figure}
\begin{lstlisting}[language=Python]
_, (w_chain, logits_chain, is_accepted_chain) = fun_mc.trace(
  (fun_mc.hamiltonian_monte_carlo_init(w_init, target), jax.random.PRNGKey(0)),
  lambda *state: fun_mc.trace(state, kernel, num_substeps, trace_mask=False),
  num_steps // num_substeps)
\end{lstlisting}
\caption{Thinning an MCMC chain. The difference from
  \Cref{fig:hmc_example} is wrapping \lstinline{kernel} in another
  \lstinline{fun_mc.trace} without collecting the intermediate
  states.}
\label{fig:thinning}
\end{figure}

\section{Reparameterization of potential functions}
\label{sec:reparameterization}

HMC is well known to be sensitive to the geometry of the target density. One
way to ameliorate this issue is through reparameterization
\citep{papaspiliopoulos2007general, parno2014transport}. In FunMC,
this is effected via function
composition, taking care to remap arguments appropriately using the
 \lstinline{fun_mc.reparam_potential_fn} utility function. It has type
\begin{lstlisting}[language=Python]
reparam_potential_fn :: (State -> (Tensor, Extra_1),
                         ReparamState -> (State, Extra_2),
                         State) ->
    (ReparamState -> (Tensor, (State, Extra_1, Extra_2)), ReparamState).
\end{lstlisting}
It accepts a density function (operating in some ``original'' space),
a diffeomorphism from the reparameterized space to the original space, and a point
in the original space.  It returns the corresponding density function in the
reparameterized space and the corresponding point in the reparameterized space.
This can be used to initialize an inference algorithm that then operates in the
reparameterized space.

While reparameterization can be advantageous for running the Markov
chain, the user may be interested in inspecting the chain's states in
the original parameterization.  FunMC's pervasive support for side
returns makes this easy to arrange: the reparameterized density
function just returns the point in the original space on the side.
The \lstinline{fun_mc.hamiltonian_monte_carlo_step} kernel propagates it, and
the kernel we wrote in \Cref{fig:reparam} extracts it on
\Cref{line:orig-state} and exposes just the original parameterization
to tracing by \lstinline{fun_mc.trace} (\Cref{line:reparam-return}).

The diffeomorphism must be an invertible function with tractable
Jacobian log-determinant.  FunMC relies on its backend to compute the
inverse and the log-determinant of the Jacobian.  One practical way to do
this is to code the diffeomorphism using the \lstinline{tfp.bijectors}
library \citep{dillon2017tensorflow} (FunMC knows the inversion and
Jacobian computation API defined thereby). The automatic inversion
mechanism being developed for JAX \citep{vikram2020probabilistic} looks like a
promising future alternative.

In summary, to reparameterize the target in our running example from
\Cref{fig:hmc_example}, we replace the kernel definition (\Cref{line:kernel})
with \Cref{fig:reparam}. Since \lstinline{fun_mc.reparameterize_potential_fn}
is just a function, the user can call it inside their kernel, using a different
diffeomorphism at each point in the chain.  This supports adapting or inferring
the diffeomorphism's parameters.

\begin{figure}[!tb]
\begin{lstlisting}[language=Python, escapechar=|]
reparam_potential_fn, reparam_w_init = fun_mc.reparameterize_potential_fn(
    target, diffeomorphism, w_init)

def kernel(hmc_state, key):
  hmc_key, key = jax.random.split(key)
  hmc_state, hmc_extra = fun_mc.hamiltonian_monte_carlo_step(hmc_state,
    reparam_potential_fn, step_size, num_integrator_steps, seed=hmc_key)
  w, logits = hmc_state.state_extra[:2] |\label{line:orig-state}|
  return (hmc_state, key), (w, logits, hmc_extra.is_accepted) |\label{line:reparam-return}|

_, (w_chain, logits_chain, is_accepted_chain) = fun_mc.trace(
    (fun_mc.hamiltonian_monte_carlo_init(reparam_w_init, reparam_potential_fn),
     jax.random.PRNGKey(0)), kernel, num_steps)
\end{lstlisting}
\caption{Reparameterization is just a matter of wrapping the original
  target.  Values in the original space can be propagated via the side
  returns instead of being lost or having to be recomputed.}
\label{fig:reparam}
\end{figure}

\section{Optimization}
\label{sec:optimization}

\begin{figure}
\begin{lstlisting}[language=Python, escapechar=|]
def kernel(hmc_state, log_step_size_state, key):
  hmc_key, key = jax.random.split(key)
  step_size = np.exp(log_step_size_state.state)
  hmc_state, hmc_extra = fun_mc.hamiltonian_monte_carlo_step(hmc_state,
    target, step_size, num_integrator_steps, seed=hmc_key)
  p_accept = np.exp(np.minimum(0., hmc_extra.log_accept_ratio))
  loss_fn = fun_mc.make_surrogate_loss_fn(lambda _: (0.8 - p_accept, ())) |\label{line:surrogate}|
  log_step_size_state, _ = fun_mc.adam_step(log_step_size_state, loss_fn,
    learning_rate=1e-2)
  w = hmc_state.state; logits = hmc_state.state_extra
  return (hmc_state, log_step_size_state, key), (w, logits, hmc_extra.is_accepted)
\end{lstlisting}
\caption{HMC with step size adaptation, reusing the FunMC Adam optimizer.
  Note the surrogate loss function trick on \Cref{line:surrogate}.}
\label{fig:adaptation}
\end{figure}

Optimization can often be cast as a Markov chain, so FunMC provides a
number of transition kernels that implement common optimization
algorithms such as gradient descent and Adam. Besides their standard
uses, these kernels are reusable for hyper-parameter adaptation in
MCMC. For example, it is conventional to tune the step size parameter
of HMC to hit an acceptance rate between 0.6 and 0.8 \citep{betancourt2014optimizing}. Given
$\alpha_t$, the acceptance rate as step $t$, the statistic $H_t = 0.8
- \alpha_t$ can be used as the gradient of a surrogate loss
function of the step size.  The step size is then adapted during the
run using some gradient-based optimization scheme such as Nesterov
dual averaging \citep{hoffman2011no}, a choice usually hardcoded into a MCMC library.

The composability of FunMC's components lets us plug in other methods instead.
For example, to use Adam, we just replace the kernel definition in
\Cref{fig:hmc_example}, \Cref{line:kernel} with \Cref{fig:adaptation}.
Since Adam expects a differentiable loss function rather than a gradient,
we need a little jiu jitsu on
\Cref{fig:adaptation}, \Cref{line:surrogate}, in the form of \lstinline{fun_mc.make_surrogate_loss_fn}.

\section{Streaming statistics}
\label{sec:streaming-statistics}

In many cases, MCMC is used to generate samples that are then summarized and
the samples discarded. Materializing all the samples is an inefficient use of memory, so FunMC
provides a number of streaming statistics to compute simple and exponentially
weighted moving means, variances and covariances. As FunMC is explicitly
batch-aware, it can compute independent statistics for each chain, but
also perform pre-aggregation across the chains.

One perennial defect of MCMC chains is poor mixing. Diagnostics to detect that
have been proposed, but standard implementations of such diagnostics require
access to the original chain history, which negates the benefit of streaming
statistics.
FunMC provides a streaming version of the Gelman potential scale
reduction $\hat{R}$ \citep{gelman1992}, a simple wrapper around FunMC's streaming variance estimator. A streaming estimator for the auto-covariance function is also provided which can be used to compute the effective sample size.

We can code a chain tracking a streaming mean, covariance, and $\hat
R$ by replacing the kernel definition \Cref{fig:hmc_example},
\Cref{line:kernel} with \Cref{fig:streaming}.
Note that $\hat{R}$ requires multiple chains. In this example, we get
them by adding a leading dimension to the chain state which indexes
independent chains.  Following the convention established by the
\lstinline{tfp.mcmc} library, FunMC interprets this encoding
by running the HMC leapfrog and Metropolis-Hastings steps on all the chains in parallel.

\begin{figure}[!tb]
\begin{lstlisting}[language=Python]
def kernel(hmc_state, cov_state, rhat_state, key):
  hmc_key, key = jax.random.split(key)
  hmc_state, hmc_extra = fun_mc.hamiltonian_monte_carlo_step(hmc_state,
    target, step_size, num_integrator_steps, seed=hmc_key)
  w, logits = hmc_state.state, hmc_state.state_extra
  cov_state, _ = fun_mc.running_covariance_step(cov_state, (w, logits), axis=0)
  rhat_state, _ = fun_mc.potential_scale_reduction_step(rhat_state, w)
  return (hmc_state, cov_state, rhat_state, key), ()

(_, fin_cov_state, fin_mean_accept_state, fin_rhat_state, _), _ = fun_mc.trace(
  (fun_mc.hamiltonian_monte_carlo_init(w_init, target),
   fun_mc.running_covariance_init((w_init.shape[-1:], y.shape), (np.float32,) * 2,
   fun_mc.potential_scale_reduction_init(w_init.shape, np.float32),
   jax.random.PRNGKey(0)), kernel, num_steps)
w_cov, logits_cov = fin_cov_state.covariance
r_hat = fun_mc.potential_scale_reduction_extract(fin_rhat_state)
\end{lstlisting}
\caption{HMC with streaming mean, covariance, and potential scale
  reduction ($\hat R$) estimation.}
\label{fig:streaming}
\end{figure}

\section{Discussion}

We have presented FunMC and illustrated multiple ways its components
can be combined to create custom MCMC and optimization algorithms.
This composition is a consequence of the core principles of
statelessness, returning side information from functions, and writing
higher-order functions to propagate this side information. We hope
this library will accelerate methodological research, either through
direct use or as an example of how to structure a composable API for
Markovian algorithms.

\begin{acks}

Authors would like to thank Matthew D. Hoffman, Jacob Burnim, Rif A. Saurous,
Dave Moore and the rest of the TensorFlow Probability team for helpful
comments.

\end{acks}

\bibliography{main}
\bibliographystyle{icml2019}

\newpage
\appendix
\section{FunMC's Hamiltonian Monte Carlo}
\label{sec:appendix}

\Cref{fig:provided_hmc} lists the pre-canned Hamiltonian Monte Carlo
in FunMC.  This is a simple composition of reusable components, which
a user can recombine to implement many HMC variants.

\begin{figure}
\begin{lstlisting}[language=Python]
def hamiltonian_monte_carlo_step(hmc_state, target_log_prob_fn, step_size,
    num_integrator_steps, seed):
  # Define the sub-transition kernels.
  kinetic_energy_fn = fun_mc.make_gaussian_kinetic_energy_fn(
    len(hmc_state.target_log_prob.shape))

  momentum_sample_fn = lambda key: fun_mc.gaussian_momentum_sample(
    state=hmc_state.state, seed=key)

  integrator_step_fn = lambda state: fun_mc.leapfrog_step(state, step_size,
    target_log_prob_fn, kinetic_energy_fn)

  integrator_fn = lambda state: fun_mc.hamiltonian_integrator(state,
    num_integrator_steps, integrator_step_fn, kinetic_energy_fn)

  # Run the integration.
  mh_key, sample_key = jax.random.split(key, 2)
  momentum = momentum_sample_fn(sample_key)

  integrator_state = fun_mc.IntegratorState(hmc_state.state,
    hmc_state.state_extra, hmc_state.state_grads, hmc_state.target_log_prob,
    momentum)

  integrator_state, integrator_extra = integrator_fn(integrator_state)

  # Do the MH accept-reject step.
  proposed_state = fun_mc.HamiltonianMonteCarloState(
    state=integrator_state.state,
    state_grads=integrator_state.state_grads,
    target_log_prob=integrator_state.target_log_prob,
    state_extra=integrator_state.state_extra)

  hmc_state, mh_extra = fun_mc.metropolis_hastings_step(hmc_state,
    proposed_state, integrator_extra.energy_change, seed=mh_key)

  return hmc_state, fun_mc.HamiltonianMonteCarloExtra(
      is_accepted=mh_extra.is_accepted,
      proposed_hmc_state=proposed_state,
      log_accept_ratio=-integrator_extra.energy_change,
      integrator_state=integrator_state, integrator_extra=integrator_extra,
      initial_momentum=momentum)
\end{lstlisting}
\caption{Default Hamiltonian Monte Carlo in FunMC.  Side quantities of
  interest are returned in the \lstinline{HamiltonianMonteCarloExtra}
  structure for possible accumulation and/or inspection by the user.}
\label{fig:provided_hmc}
\end{figure}
\end{document}